# A broadband achromatic polarization-insensitive metalens consisting of anisotropic nanostructures

Wei Ting Chen[1,*], Alexander Y. Zhu[1], Jared Sisler[1,2], Zameer Bharwani[1,2] and Federico Capasso[1,*]

[1]Harvard John A. Paulson School of Engineering and Applied Sciences, Harvard University, Cambridge, Massachusetts 02138, USA

[2]University of Waterloo, Waterloo, ON N2L 3G1, Canada

*Corresponding author: capasso@seas.harvard.edu and weitingchen@seas.harvard.edu

Metasurfaces have attracted widespread attention due to an increasing demand of compact and wearable optical devices. For many applications, polarization-insensitive metasurfaces are highly desirable and appear to limit the choice of their constituent elements to isotropic nanostructures. This greatly restricts the degrees of geometric parameters available in designing each nanostructure. Here, we demonstrate a polarization-insensitive metalens using otherwise anisotropic nanofins which offer additional control over the dispersion and phase of the output light. As a result, we can render a metalens achromatic and polarization-insensitive across nearly the entire visible spectrum from wavelength $\lambda$ = 460 nm to 700 nm, while maintaining diffraction-limited performance. The metalens is comprised of just a single layer of $TiO_2$ nanofins and has a numerical aperture of 0.2 with a diameter of 26.4 µm. The generality of our polarization-insensitive design allows it to be implemented in a plethora of other metasurface devices with applications ranging from imaging to virtual/augmented reality.

**Introduction**

Metasurfaces, comprising sub-wavelength spaced nanostructures at an interface, provide the means to accurately control the properties of light, including phase, amplitude, and polarization [1-4]. This allows for the possibility of highly compact and efficient devices [5-11]. Amongst these devices, metalenses have attracted intense interest due to their applicability to both consumer (phone cameras, virtual/augmented reality headsets) and industry products (microscopy, lithography, sensors, and displays) [12-20]. Recent works have focused on developing the broadband achromatic focusing capabilities of metalenses in the visible spectrum [21, 22]. However, these metalenses suffer from polarization sensitivity, i.e., they can only focus light with a certain circular polarization. This challenge can be overcome by using symmetric cylindrical or square-shaped nano-pillars [23-25]. However, by doing so we lose a degree of freedom in the design space due to the symmetry. Here, somewhat counterintuitively, we show that it is indeed possible to simultaneously achieve an achromatic and polarization-insensitive metalens in the visible using *anisotropic* $TiO_2$ nanofins, a different solution compared to recent publications associated with spatial multiplexing and symmetry [26, 27]. These anisotropic nanofins allow us to accurately implement the phase and its higher-order derivatives (i.e. group delay and group delay dispersion) with respect to frequency simultaneously. We designed and fabricated a metalens with a numerical aperture (NA) of 0.2. The metalens exhibits a measured focal length shift of only 9 % and has diffraction-limited focal spots from wavelength λ = 460 to 700 nm. The focusing efficiency of the metalens varies by only ~ 4% under various incident polarizations. To showcase the generality of our principle, we also demonstrate a polarization-insensitive metasurface with diffraction efficiency of about 92% at wavelength $\lambda$ = 530 nm.

**Principle**

To achromatically focus a broadband incident beam in a diffraction limited spot, a metalens must impart a spatial- and frequency-dependent phase profile given by

$$\varphi(\mathrm{r},\omega) = -\frac{\omega}{c}(\sqrt{r^2 + F^2} - F) \quad (1)$$

where $r$, $\omega$, and $F$ are the lens radial coordinate, angular frequency, and a constant focal length, respectively. The Taylor expansion of Eq. 1:

$$\varphi(\mathrm{r},\omega) = \varphi(\mathrm{r},\omega_d) + \left.\frac{\partial \varphi}{\partial \omega}\right|_{\omega=\omega_d}(\omega-\omega_d) + \left.\frac{\partial^2 \varphi}{2\partial \omega^2}\right|_{\omega=\omega_d}(\omega-\omega_d)^2 + \ldots \quad (2)$$

identifies the required phase $\varphi(\mathrm{r},\omega_d)$, group delay $\left.\frac{\partial \varphi}{\partial \omega}\right|_{\omega=\omega_d}$, and group delay dispersion $\left.\frac{\partial^2 \varphi}{\partial \omega^2}\right|_{\omega=\omega_d}$ that needs to be fulfilled at every lens coordinate $r$. An intuitive way to understand each term in Eq. 2 is to treat the incident light as wavepackets. The required phase profile sends incident wavepackets towards the focus, while the first and the higher order derivative terms ensure that the incident wavepackets arrive at the focus simultaneously and identically in the time domain, respectively [21]. The challenge here lies in the fact that the chosen nanostructures must satisfy each derivative term in Eq. 2 at every lens coordinate. Previous designs made use of the geometric (or Pancharatnam-Berry) phase principle to decouple the phase $\varphi(\mathrm{r}, \omega_d)$ from the dispersion (group delay and group delay dispersion) [21, 22, 28, 29]. However, this approach also comes with an unwanted polarization-sensitivity, i.e. these metalenses can only focus incident light with a particular circular polarization.

Our design principle still involves Pancharatnam-Berry phase; however, we circumvent the aforementioned drawback by limiting the rotation angle of each anisotropic element to either 0 or 90 degrees. Each element is comprised of multiple nanofins to provide additional degrees of freedom to engineer the dispersion (Fig. 1(a), inset). The layout of a fraction of our achromatic and polarization-insensitive metalens is depicted in Fig. 1(a) and a scanning electron microscope image of a region of our fabricated metalens is shown in Fig. 1(b). To tune the phase and dispersion, each nanofin's length and width is varied and the gap $g$ between nanofins is set to be either 60 nm or 90 nm. By using anisotropic elements instead of the previously standard symmetrical circular or square pillars [23-25], we have more geometric parameters to alter for better dispersion control. More importantly, the anisotropic elements offer a freedom to impart an additional $\pi$ phase shift without changing their dispersion characteristics. This is essential in order to fulfill both the required phase and dispersion given by Eq. 2, and can be explained by the Pancharatnam-Berry phase [30, 31]. When light passes through a nanofin, the transmitted electric field can be described by the Jones vector [32]:

$$\begin{bmatrix} \tilde{E}_x \\ \tilde{E}_y \end{bmatrix} = \frac{\tilde{t}_l + \tilde{t}_s}{2} \begin{bmatrix} 1 \\ \pm i \end{bmatrix} + \frac{\tilde{t}_l - \tilde{t}_s}{2} \exp^{\pm i 2\alpha} \begin{bmatrix} 1 \\ \mp i \end{bmatrix} \quad (3)$$

$\tilde{t}_l$ and $\tilde{t}_s$ represent complex transmission coefficients when the normalized electric field of the incident light is polarized along the long and short axis of the nanofin, respectively. The $\alpha$ term is defined as the counterclockwise rotation angle of the nanofin with respect to the *x*-axis. The first term of Eq. 3 causes unwanted scattering and can be minimized if the nanofin is designed as a miniature half-waveplate. In this case, the amplitude of the second term $abs(\frac{\tilde{t}_l - \tilde{t}_s}{2})$ increases, corresponding to maximal polarization conversion efficiency. The $\exp^{\pm i 2\alpha}$ in the second term is accompanied by a polarization converted term and illustrates the origin of Pancharatnam-Berry phase. Under left-handed circularly polarized incidence, a rotation of $\alpha$ imparts a frequency-independent phase of $2\alpha$ to the right-handed circularly polarized output light without affecting the dispersion, which is determined by $\frac{\tilde{t}_l - \tilde{t}_s}{2}$. This usually results in polarization-sensitivity because the values of $\exp^{i 2\alpha}$ and $\exp^{-i2\alpha}$, obtained under left and right circular polarized (LCP and RCP) incident light, respectively, are not identical. However, if one arranges the nanofin with $\alpha = 0°$ or $\alpha = 90°$, their values become equal. Therefore, both RCP and LCP incident light will experience the same phase profile upon interacting with a metasurface consisting of either mutually parallel or perpendicular nanofins. Since any incident polarization can be decomposed into a combination of LCP and RCP, this property implies that the metasurface is polarization insensitive. Figure 1(c) confirms the results predicted by Eq. 3. A metalens element provides the same phase for RCP (line) and LCP (circles) incidence, and, for a given circular polarization incidence, a 90 degree rotation imparts a $\pi$ phase shift without affecting group delay (slope) and group delay dispersion (curvature).

**Design of an achromatic and polarization-insensitive metalens**

The design of our polarization-insensitive and achromatic metalens starts from a parameter sweep of the element shown in the inset of Fig. 1(a) to build a library. We used a finite-difference time-domain (FDTD) solver to obtain each element's phase at $\lambda = 530$ nm, as well as its group delay and group delay dispersion. More simulation details can be found in our previous publication [21]. Figure 2(a) plots the three quantities of interest, phase, group delay, and group delay dispersion, at the design wavelength of 530 nm for each element. There are over ten-thousand geometrical combinations, resulting in a dense scatter plot of dots to fine tune the dispersion. Note that due to the principle outlined in Fig. 1(c), an element rotated by 90 degrees (i.e. purple points) will experience a $\pi$ phase shift for all frequencies with no change in the values of its dispersion. As a result, the library can be extended, allowing for better implementation of the required phase and dispersion (black symbols), which were calculated based on Eq. 1 for an achromatic metalens with a diameter of 26.4 μm and NA = 0.2. To realize the metalens, the elements selected must be those closest to the required (black) points in the 3-dimensional space of phase, group delay, and group delay dispersion displayed in Fig. 2(a). Because only the relative values of these parameters are important, the library can be shifted in this 3-dimensional space to better fit the required values. A particle swarm optimization method was used to find the optimal shifts for phase, group delay, and group delay dispersion, which minimizes the distance between each required point and the values provided by the elements in our library. The final results can be better visualized in Figs. 2(b)-Fig. 2(d). The phase, group delay, and group delay dispersion of the selected metalens elements are shown in blue, together with the corresponding required values (black curves). We only consider terms up to the group delay dispersion because the values of any higher orders for our selected elements are very small.

**Results and Discussion**

We subsequently fabricated the achromatic and polarization insensitive metalens using electron beam lithography followed by atomic layer deposition of $TiO_2$ and resist removal [33] and compared its performance to a diffractive metalens of the same diameter and NA. The diffractive metalens was designed using a nanofin with the same length and width, but varying rotation angle. The diffractive metalens represents the case without dispersion engineering and has a focal length shift similar to a Fresnel lens. We also show in Supplementary Video 1 simulation results for a complete metalens with a smaller lens diameter and a higher NA of 0.6 and confirm its achromatic and polarization-insensitive focusing behavior. The focal length shifts of the fabricated achromatic and diffractive metalenses were determined by measuring their point spread functions at each wavelength along the propagation direction (*z*-axis) with 1 μm resolution (Fig. 3(a)). The left panel in Fig. 3(a) demonstrates a small focal length variation of about 6 μm for the achromatic metalens compared to that of 30 μm in the diffractive metalens (right panel). The normalized intensity profiles along the white dashed lines can be seen in Fig. 3(b), with the top panel corresponding to the achromatic metalens and the bottom panel corresponding to the chromatic metalens. The achromatic metalens is diffraction-limited and its focal sizes and Strehl ratios as a function of wavelengths are given in Fig. S1. The achromatic metalens was also shown to be polarization insensitive by measuring the efficiency of the focal spot with changing polarizations of incident light. This measurement was performed by changing the angle of linearly polarized incident light from 0º to 90° in steps of 4°. The measured efficiencies weakly change with polarization, as shown in Fig 3(c), which constitutes proof of polarization-insensitive focusing. The green and blue symbols show the measured efficiencies at a single wavelength of 532 nm and under broadband light from 470-670 nm, respectively. Note that the polarization state of the metalens' focal spot can be different from that of the incident light due to the phase of the polarization converted term (the 2$^{nd}$ term of Eq. 3).

It is worth noting that the metalens efficiency shown in Fig. 3(b) is lower than our previous chromatic metalenses [17, 34]. This can be explained by the fact that some low polarization conversion elements were selected to cover a large range of dispersion values for achromaticity (see Fig. S2 for a plot of efficiency and dispersion). However, we emphasize that our polarization-insensitive design approach does not exclude the design of highly-efficient metasurfaces. For example, we show in Fig. 4(a) the layout of a conventional chromatic metasurface beam deflector designed for wavelength $\lambda$ = 530 nm with an output diffraction angle of $\theta$ = 15º. Figure 4(b) shows the normalized far-field power across the visible under *x*-polarized incidence as a function of wavelength. The metasurface has mainly a single diffraction beam over a bandwidth of 50 nm centered at 530 nm and results in a high diffraction efficiency of about 92%. The diffraction efficiency is defined as the power of the first (+1) diffraction order divided by that of transmitted power. We numerically verified in Fig. 4(c) that such a high diffraction efficiency is maintained under various linearly and circularly polarized incident beams. It can be seen that at a given wavelength, the efficiency remains relatively constant across all polarizations, highlighting the polarization insensitivity. The absolute efficiency at $\lambda = 530$ nm, i.e. the power diffracted to 15 degrees divided by total incident power, is about 70% (see Fig. S3 for a plot of the absolute efficiency of the metasurface).

**Conclusions**

We have demonstrated with simulations and experiments a general principle for designing polarization-insensitive metasurfaces using anisotropic nanostructures as building blocks. These anisotropic structures allow for a more accurate implementation of phase, group delay, and group delay dispersion, while simultaneously making it possible to realize a polarization-insensitive, diffraction-limited and achromatic metalens from wavelength $\lambda = 460$ nm to 700 nm. Our design approach of polarization-insensitivity is also valid for other metasurface devices with applications in imaging and augmented reality.

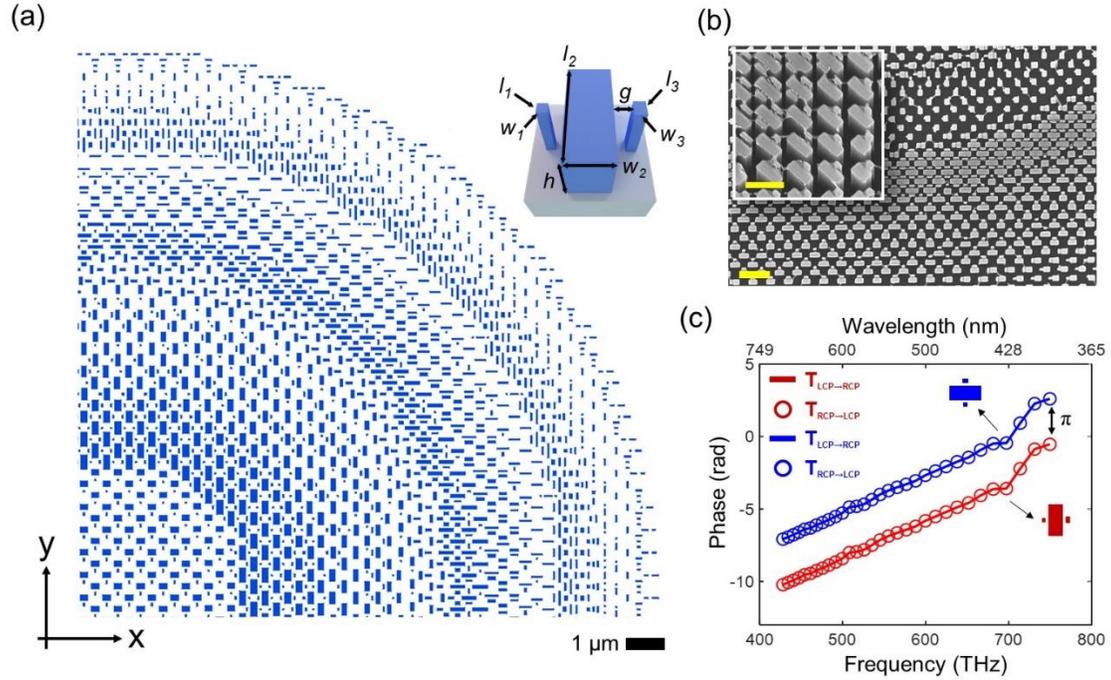

**Figure 1: Principle behind the polarization-insensitive and achromatic metalens.** (a) Layout of a quadrant of the metalens. It has an NA of 0.2 and a diameter of 26.4 μm. The inset shows a schematic diagram of its constituent elements. Each element comprises $TiO_2$ nanofins of the same height $h = 600$ nm. These metalens elements are spaced equally along the *x*- and *y*- directions with a lattice constant of 400 nm. (b) A scanning electron microscope image of a part of the fabricated metalens. Scale bar: 1 μm. The inset shows a magnified and oblique view of the nanofins. Scale bar: 500 nm. (c) Simulated phase shift of the component of the transmitted electric field with polarization orthogonal to the incident circularly polarized light. The legend, for example $T_{LCP \rightarrow RCP}$, represents the phase of RCP transmitted light under LCP incidence. The blue and red colors show the same

element, consisting of three nanofins, oriented along horizontal and vertical directions, respectively. The nanofin parameters ($w_1, l_1, w_2, l_2, w_3, l_3, g$) = (50, 50, 170, 370, 50, 90, 60) in nanometer units. The element shows identical phase under both RCP and LCP illuminations. Note that for a given circular polarization, a 90-degree rotation introduces a $\pi$ phase shift without affecting group delay (slope) and group delay dispersion (curvature).

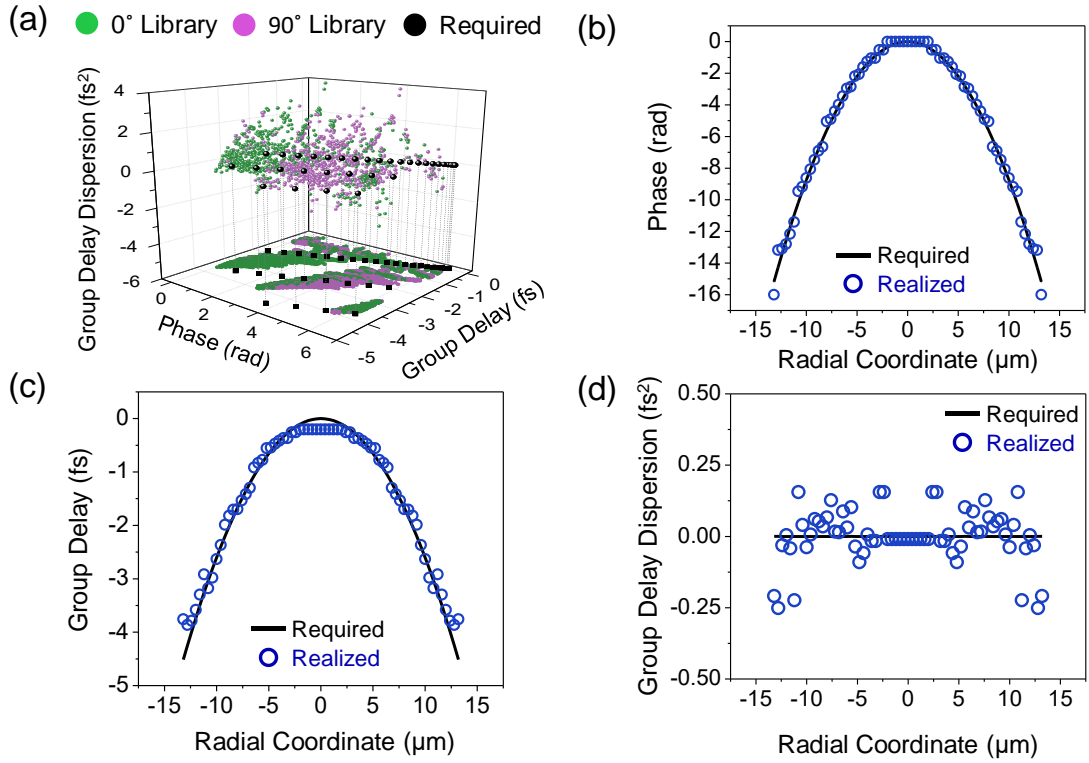

**Figure 2: A comparison of required and realized phase, group delay, and group delay dispersion for the metalens shown in Fig. 1(a).** (a) Phase, group delay, and group delay dispersion for all elements in our simulation library (colored points) and required values (black points). Each element is represented by a green and purple point in the plot because a 90-degree rotation can impart a phase change of $\pi$ without changing its group delay and group delay dispersion. (b)-(d) Realized (blue circles) and required (black curves) phase, group delay, and group delay dispersion at each radial coordinate across the polarization-insensitive and achromatic metalens.

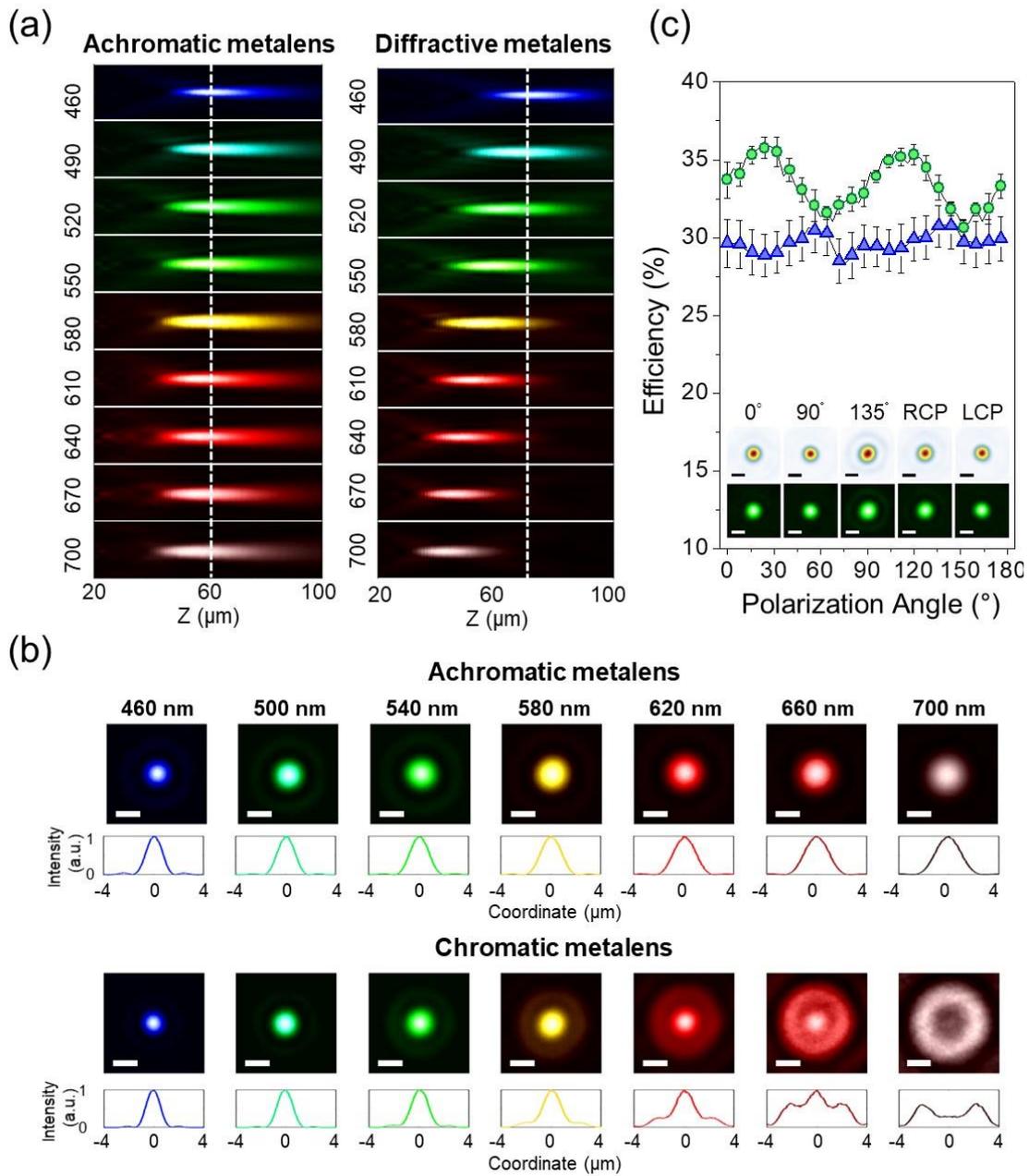

**Figure 3: Measured focal spot profiles (in false colors) and efficiencies for achromatic and diffractive metalenses.** The measured metalenses have a NA = 0.2 and a focal length of 67 μm at $\lambda$ = 560 nm. (a) Measured intensity distributions in the *y-z* plane shown in false colors corresponding to their respective wavelengths in the visible (labelled to left of plots

in nanometers). The left and right panels correspond to achromatic and diffractive metalenses respectively. The latter, as a control sample, was realized without dispersion engineering and has a focal length shift similar to that of Fresnel lenses. Incident light travels along the positive *z*-axis. (b) Normalized intensity profiles along the white dashed lines of (a), for the achromatic metalens (top) and the diffractive metalens (bottom). The position of the dashed line corresponds to the focal length at λ= 460 nm. (c) Efficiency of the achromatic metalens as a function of the angle of linearly polarized incident light in steps of 4º. The illumination light sources are alternately a single wavelength 532 nm diode laser and a tunable broadband laser with 200 nm bandwidth centered at 570 nm. The measured efficiencies using the monochromatic and broadband light source are represented by the green and blue symbols, respectively. The inset shows the focal spot profile, with the top and bottom rows corresponding to the diode (monochromatic) and tunable broadband laser illumination, respectively. The polarizations of input light are labelled at the top. Scale bars: 2 μm.

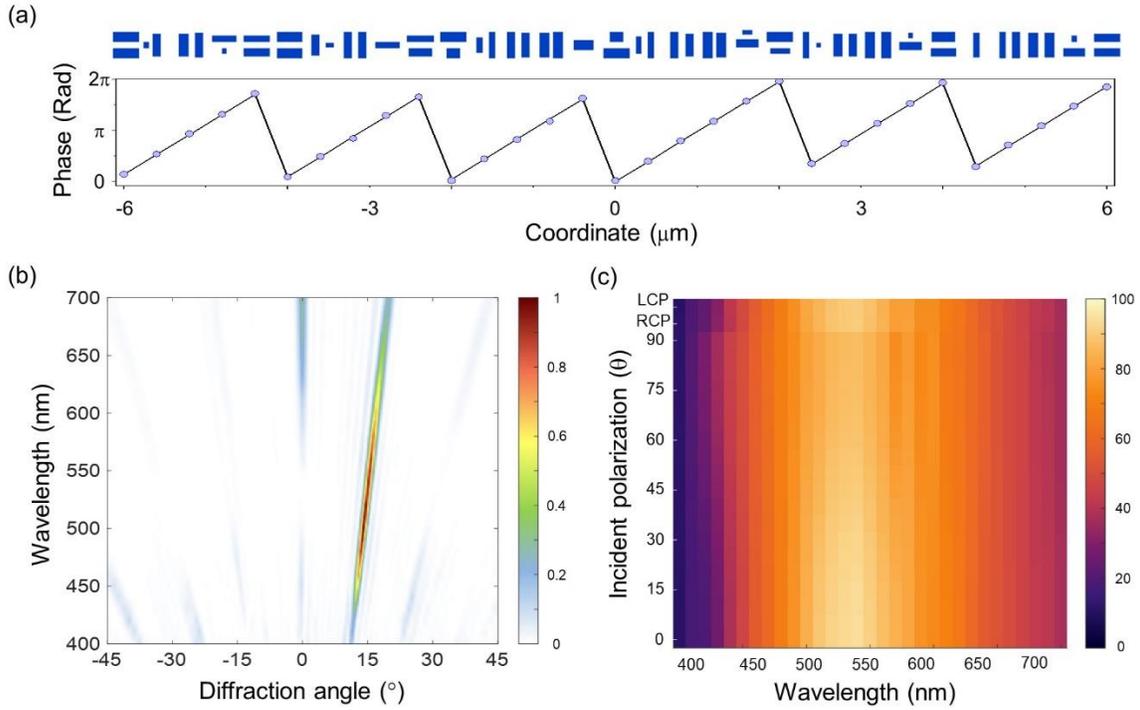

**Figure 4: Simulated efficiency for a polarization-insensitive phase-gradient metasurface.** (a) Layout of the designed metasurface. The metasurface is comprised of mutually parallel and perpendicular nanofins with the geometries and orientations chosen to deflect a normal incident beam to an angle of 15 degrees at the design wavelength of 530 nm. The bottom panel shows the target and realized phases in a black line and blue circles, respectively. (b) Normalized far-field power under x-polarized incidence as a function of incident wavelength and diffraction angles. (c) Diffraction efficiency for the metasurface across the visible spectrum under linear and circular polarizations. The polarization angles are labelled on the y-axis, while the last two rows showing the cases

for right- and left-handed polarizations. At each wavelength, the efficiency is maintained at a relatively constant value, thus displaying the polarization insensitivity.

**Supplementary Materials:**

Figure S1-S3

Supplementary Movie

**Author contributions:**

W. T. C. and F.C. conceived the study. A. Y. Z. fabricated the samples. W. T. C., J. S. and Z. B. performed simulations and developed codes. W. T. C., A. Y. Z. and J. S. measured the hybrid lens. All authors wrote the manuscript, discussed the results, and commented on the manuscript.

**Competing financial interests**

The authors declare no competing financial interests

**Acknowledgements**

This work was supported by the Air Force Office of Scientific Research (MURI, grant# FA9550-14-1-0389 and grant# FA9550-16-1-0156) and the Defense Advanced Research Projects Agency (grant# HR00111810001). This work was performed in part at the Center for Nanoscale Systems (CNS), a member of the National Nanotechnology Coordinated Infrastructure Network (NNCI), which is supported by the National Science Foundation under NSF award no. 1541959. Federico Capasso gratefully acknowledges a gift from Huawei Inc. under its HIRP FLAGSHIP program.**References:**

1. V.-C. Su, C. H. Chu, G. Sun, and D. P. Tsai, "Advances in optical metasurfaces: fabrication and applications," *Opt. Express* **26**, 13148-13182 (2018).
2. A. V. Kildishev, A. Boltasseva, and V. M. Shalaev, "Planar photonics with metasurfaces," *Science* **339**, 1232009 (2013).
3. N. Yu, P. Genevet, M. A. Kats, F. Aieta, J.-P. Tetienne, F. Capasso, and Z. Gaburro, "Light Propagation with Phase Discontinuities: Generalized Laws of Reflection and Refraction," *Science* **334**, 333-337 (2011).
4. M. Qiu, M. Jia, S. Ma, S. Sun, Q. He, and L. Zhou, "Angular Dispersions in Terahertz Metasurfaces: Physics and Applications," *Phys. Rev. A* **9**, 054050 (2018).
5. A. Y. Zhu, W.-T. Chen, M. Khorasaninejad, J. Oh, A. Zaidi, I. Mishra, R. C. Devlin, and F. Capasso, "Ultra-compact visible chiral spectrometer with meta-lenses," *APL Photonics* **2**, 036103 (2017).
6. N. A. Rubin, A. Zaidi, M. Juhl, R. P. Li, J. B. Mueller, R. C. Devlin, K. Leósson, and F. Capasso, "Polarization state generation and measurement with a single metasurface," *Opt. Express* **26**, 21455-21478 (2018).
7. N. I. Zheludev, "Obtaining optical properties on demand," *Science* **348**, 973-974 (2015).
8. A. Pors, M. G. Nielsen, and S. I. Bozhevolnyi, "Plasmonic metagratings for simultaneous determination of Stokes parameters," *Optica* **2**, 716-723 (2015).
9. G. Zheng, H. Mühlenbernd, M. Kenney, G. Li, T. Zentgraf, and S. Zhang, "Metasurface holograms reaching 80% efficiency," *Nat. Nanotechnol.* **10**, 308-312 (2015).
10. K. Huang, F. Qin, H. Liu, H. Ye, C.-W. Qiu, M. Hong, B. Luk'yanchuk, and J. Teng, "Planar Diffractive Lenses: Fundamentals, Functionalities, and Applications," *Advanced Materials* **30**, 1704556 (2018).
11. S. Colburn, A. Zhan, and A. Majumdar, "Metasurface optics for full-color computational imaging," *Sci. Adv.* **4**(2018).
12. C. Schlickriede, N. Waterman, B. Reineke, P. Georgi, G. Li, S. Zhang, and T. Zentgraf, "Imaging through nonlinear metalens using second harmonic generation," *Advanced Materials* **30**, 1703843 (2018).
19